\documentclass[]{spie}  

\usepackage{amsmath,amsfonts,amssymb}
\usepackage{siunitx}
\usepackage{graphicx}
\usepackage[colorlinks=true, allcolors=blue]{hyperref}
\usepackage{subcaption}
\usepackage{float}

\title{SPECULOOS: five years hunting terrestrial planets around ultra-cool dwarfs}

\author[a]{Sebastián Zúñiga-Fernández}
\author[a]{Michaël Gillon}
\author[c,d]{Peter P. Pedersen}
\author[e]{Daniel Sebastian}
\author[f]{Artem Burdanov}
\author[g]{Brice-Olivier Demory}
\author[e]{Amaury H. M. J. Triaud}
\author[f]{Julien de Wit}
\author[b]{Emmanuël Jehin}
\author[c,d]{Didier Queloz}
\author[h]{Lionel J. Garcia}
\author[a]{Mathilde Timmermans}
\author[a,f,i]{Khalid Barkaoui}
\author[e]{Thomas Baycroft}
\author[e]{Yasmin Davis}
\author[a]{Fatemeh Davoudi}
\author[a]{Laetitia Delrez}
\author[j]{Elsa Ducrot}
\author[n]{Yilen G\'omez Maqueo Chew}
\author[c]{Matthew J. Hooton}
\author[c]{Clàudia Janó-Muñoz}
\author[f,k]{Benjamin V. Rackham}
\author[e]{Madison Scott}
\author[c]{Samantha Thompson}
\author[l,m,a]{Selçuk Yal\c{c}inkaya}

\affil[a]{Astrobiology Research Unit, Universit\'e de Li\`ege, Liège, Belgium}
\affil[b]{Space Sciences, Technologies and Astrophysics Research (STAR) Institute, Universit\'e de Li\`ege, Liège, Belgium}
\affil[c]{Cavendish Laboratory, JJ Thomson Avenue, Cambridge CB3 0HE, United Kingdom}
\affil[d]{Department of Physics, ETH Zurich, Zurich, Switzerland}
\affil[e]{School of Physics \& Astronomy, University of Birmingham, Birmingham, United Kingdom}
\affil[f]{Department of Earth, Atmospheric and Planetary Science, Massachusetts Institute of Technology, Cambridge, MA, USA}
\affil[g]{Center for Space and Habitability, University of Bern, Bern, Switzerland}
\affil[h]{Center for Computational Astrophysics, Flatiron Institute, New York, NY, USA}
\affil[i]{Instituto de Astrofísica de Canarias (IAC), La Laguna, Tenerife, Spain}
\affil[j]{LESIA, Observatoire de Paris, Université PSL, CNRS, Sorbonne Université, Meudon, France}
\affil[k]{Kavli Institute for Astrophysics and Space Research, Massachusetts Institute of Technology, Cambridge, MA, USA}
\affil[l]{Ankara University, Astronomy and Space Sciences Research and Application Center (Kreiken Observatory), Ankara, T\"urkiye}
\affil[m]{Department of Astronomy \& Space Sciences, Ankara University, Ankara, T\"urkiye} 
\affil[n]{Instituto de Astronomía, Universidad Nacional Autónoma de México, Ciudad de México, México}

\authorinfo{Further author information: (Send correspondence to S. Zúñiga-Fernández and M. Gillon)\\S. Zúñiga-Fernández: E-mail: sgzuniga@uliege.be\\  M. Gillon (Project leader): E-mail: michael.gillon@uliege.be}

\begin{document} 
\maketitle

\begin{abstract}
The SPECULOOS (Search for habitable Planets EClipsing ULtra-cOOl Stars) project aims to detect temperate terrestrial planets transiting nearby ultracool dwarfs, including late M-dwarf stars and brown dwarfs, which are well-suited for atmospheric characterization using the James Webb Space Telescope (JWST) and upcoming giant telescopes like the European Extremely Large Telescope (ELT). Led by the University of Liège, SPECULOOS is conducted in partnership with the University of Cambridge, the University of Birmingham, the Massachusetts Institute of Technology, the University of Bern, and ETH Zurich. The project operates a network of robotic telescopes at two main observatories: SPECULOOS-South in Chile, with four telescopes, and SPECULOOS-North in Tenerife, currently with one telescope (soon to be two). This network is complemented by the SAINT-EX telescope located in San Pedro Mártir, Mexico. In this paper, we review the status of our facilities after five years of operations, the current challenges and development plans, and our latest scientific results.
\end{abstract}


\section{INTRODUCTION}
\label{sec:intro}  
The successful launch and operation of the James Webb Space Telescope (JWST\cite{Gardner2006}) and the ongoing construction of extremely large telescopes (ELTs) brings a new era in the studies of extra-solar planets\cite{Cowan2015,Greene2016}. These advancements enable the atmospheric characterisation of transiting temperate earth-like planets, potentially allowing the detection of chemical traces of life beyond our solar system\cite{Seager2016}. In that regard, the best target for detection of bio-signatures would be an habitable-zone terrestrial planet transiting one of the nearest ultracool dwarfs (UCDs). These very-low mass stars and brown dwarfs have characteristic that make them ideal candidates to search for transiting temperate rocky worlds and also optimal for following atmospheric characterisation\cite{Burdanov2018}. UCDs offer several advantages for studying transiting Earth-sized planets. Their Jupiter-sized results in much deeper transit signals compared to Earth-Sun twin systems, with expected transit depths ranging from 0.5\% to over 1\% for Earth-sized planets, detectable by many ground-based surveys\cite{Nutzman2008}. Additionally, UCDs have a closer habitable zone due to their low luminosity, increasing the likelihood (geometric transit probabilities higher than 1.5\%) and frequency (up to 100 transits per year) of transits of potentially habitable planets. These characteristics, combined with the large planet-to-star flux and size ratios, make UCDs ideal for atmospheric characterisation of transiting planets.

Ongoing surveys like ExTrA\cite{Bonfils2015} targeting late M-dwarfs, and PINES\cite{Tamburo2022} targeting L- and T-type dwarfs, have begun their operation but not yet reported any detection. Finished surveys like EDEN\cite{Gibbs2020}, targeting late M-dwarfs, has not found either any transiting planets detection \cite{Dietrich2023}. Currently, the only two known systems with transiting planets orbiting a UCD are TRAPPIST-1\cite{Gillon2017,Luger2017,Agol2021} (=SPECULOOS-1), which hosts seven exoplanets forming a unique near-resonant chain with three of them located in the habitable zone, and the recently discovered SPECULOOS-3b\cite{Gillon2024}. The TRAPPIST-1 system was discovered by the TRAPPIST UCD transit survey\cite{Gillon2011,Jehin2011}, which observed 50 of the brightest southern UCDs for about 100 hours each with the TRAPPIST-South telescope in Chile. The TRAPPIST survey served as a prototype for a more ambitious search for exoplanets around UCDs~-~SPECULOOS (Search for habitable Planets EClipsing ULtra-cOOl Stars). The detection – and even the very existence – of the TRAPPIST-1 planetary system fully demonstrates the instrumental concept and the scientific potential of the SPECULOOS project.

A general overview of the SPECULOOS project is presented in Refs. \citenum{Gillon2018,Burdanov2018}. Delrez et al. (2018)\cite{Delrez2018} provide technical details and early results regarding the photometric performance of the first two telescopes installed at the SPECULOOS South Observatory (SSO) in Paranal. The dedicated SPECULOOS pipeline and the performance of all SSO telescopes during the first year of operation are presented in Ref. \citenum{Murray2020}. The target list and observational strategy for the SPECULOOS programs are presented in Ref. \citenum{Sebastian2021}, while the latest developments of the project as a whole are detailed in Ref. \citenum{Sebastian2020DevelopmentProject}. In this report, we present an update on our project after five years of operation. In Section \ref{sec:obs_facilities}, we describe the observatories that form our SPECULOOS network and their instrumentation. Section \ref{sec:obs_operations} describes our robotic observatory operations, scheduling, and observing strategy. In Section \ref{sec:data_reduc}, we describe our archive, data reduction pipelines, and light curve analysis tools. In Section \ref{sec:devs}, we show our current developments and finally discuss our latest scientific results obtained with the SPECULOOS observatories in Section \ref{sec:results}.

\section{SPECULOOS Network of observatories}
\label{sec:obs_facilities}
The SPECULOOS network consists of three facilities (see Table \ref{tab:SPECULOOS_network}). Four telescopes are installed at the ESO Paranal Observatory (Atacama desert, Chile), named Io, Europa, Ganymede and Callisto. These telescopes compose the SPECULOOS South Observatory (SSO \cite{Delrez2018,Sebastian2020DevelopmentProject}, IAU code W75), which has been operational since January 2019. The second node, the SPECULOOS North Observatory (SNO \cite{Burdanov2022}, IAU code Z25), is currently composed of one telescope that is located at the Teide Observatory (Canary islands, Spain), named Artemis, and which has been operational since June 2019. By the summer of 2025, a new telescope named Orion will be installed at SNO, and which will be a twin of our other SPECULOOS telescopes with the improvement that it will be able to use two cameras (one of them an infrared camera, see Sec. \ref{sec:spirit}) at the same time using a dichroic mirror. The third node is the SAINT-EX (Search And characterIzatioN of Transiting EXoplanets \cite{Demory2020,GomezMaqueo2023}) telescope in San Pedro Mártir observatory (Mexico), operational since March 2019. Each of these observatories are devoting 70\% of their usable observational time to the SPECULOOS survey. In addition, the two 60cm TRAPPIST telescopes\cite{Jehin2011,Gillon2011,Barkaoui_2017}, while not officially part of the SPECULOOS network, allocate a fraction of their time to support the survey, focusing on its brightest targets.

Each SPECULOOS observatory consists of identical robotic Ritchey-Chretien (F/8) telescopes with a 1-meter aperture build by ASTELCO company. Each telescope is equipped with an Andor iKon-L thermoelectrically cooled camera, featuring a near-infrared optimized, deep depletion 2k x 2k e2v CCD detector (13.5 \unit{\um} pixel size). The field of view on the sky is 12 x 12 arcminutes, resulting in a pixel scale of 0.35 arcseconds/pixel. Exposure control is managed by a mechanical shutter with overlapping iris blades. Although these shutters are generally very durable, they have a limited lifespan. To extend shutter longevity, we limit our remote observations to exposure times longer than 10 seconds. The camera is typically operated at -60°C (achieved via five-stage Peltier cooling) with a dark current of 0.3 electrons/s/pixel. The detector offers high sensitivity across a wide wavelength range (350-950 nm), with maximum quantum efficiency of 94\% at both 420 and 740 nm. Each camera is equipped with its own filter wheel, providing the Sloan g’, r’, i’, z’ filters, and two special exoplanet filters: the near-infrared luminance I+z filter (with transmittance $> 90\%$ from 750 to beyond 1000 nm, primarily used for the SPECULOOS core program), and a blue-blocking filter called Exo (with transmittance $> 90\%$ from 500 to beyond 1000 nm). In March 2023 an infrared camera was installed in Callisto telescope at SSO, this camera is an InGaAs 1280x1024 CMOS-based camera. The field of view on the sky is 5.3 x 6.6 arcminutes (further details in Sec. \ref{sec:spirit}).

\begin{table}[ht]
    \caption{Observatories of the SPECULOOS network.}
    \centering
    \vspace{3pt}
    \begin{tabular}{|l|l|l|l|l|}
    \hline
    \rule[-1ex]{0pt}{3.5ex} Observatory  &  Telescope & Host Observatory & Coordinates & Height \\
    \hline
    \rule[-1ex]{0pt}{3.5ex} SSO  & 4 & ESO Paranal Observatory, Chile & 24.61596 S 70.39057 W & 2490 m \\
    \hline
    \rule[-1ex]{0pt}{3.5ex} SNO  & 1 (soon 2) & Teide Observatory, Tenerife, Spain & 28.30000 N 16.51158 W & 2438 m \\
    \hline
    \rule[-1ex]{0pt}{3.5ex} SAINT-EX  & 1 & OAN-San Pedro Mártir, Mexico & 31.04342 N 115.45476 W & 2780 m \\
    \hline
    \end{tabular}
    
    \label{tab:SPECULOOS_network}
\end{table}

All telescopes require minimum on-site maintenance. The host observatories provide emergency help in case of technical difficulties as well as regular mirror cleaning and check-ups to ensure continuous robotic operations. We plan a service mission once a year where we carry on a complete inspection of our facilities, perform general maintenance, replenish the stock of spare parts and update our inventory.

\section{Observation strategy and operations}
\label{sec:obs_operations}
The SPECULOOS observatories are robotic and can be controlled from anywhere there is internet access through a Virtual Private Network (VPN) connection. Before local sunset, the telescope operator must check the status of our telescope and the weather conditions in our cameras, satellite images and data from weather stations. If all the systems are working properly and we are not in dangerous conditions such as storms, rain or \textit{calima} (dust storm in the case of SNO), the operator can startup the night's observing plans. These plans, which consist of simple text files, are created automatically using \texttt{SPOCK}\footnote{\url{https://github.com/educrot/SPOCK}}(SPeculoos Observatory sChedule maKer \cite{Sebastian2020DevelopmentProject}) and managed by \texttt{ACP}\footnote{\url{www.acp.dc3.com}} Observatory Control Software installed on the control computer of each telescope. After the startup by the operator, ACP handles all necessary actions during the night, including opening the dome, executing science observations, performing twilight flats, dome closing after sunrise, and conducting bias and dark calibrations. In the event of a dome closure due to a weather alert during the night, \texttt{ACP} can automatically restart the observation after a safe weather time window, thanks to a custom routine\cite{Sebastian2020DevelopmentProject}.

The SPECULOOS target catalogue contains an homogeneous selected sample of close-by ($< 40$ pc) low-mass stars and UCDs. This target list was build to match SPECULOOS science goals and could be divided into three non-overlapping programs:
\begin{description}
    \item[Program 1 (365 targets)] Includes all targets that allow transit transmission spectroscopy with JWST for Earth-like planets.
    \item[Program 2 (171 targets)] Includes all targets that allow a detection of temperate Earth-size planets, like TRAPPIST-1b, with TESS.
    \item[Program 3] Includes 1121 targets with the spectral type M6 and later and aims to explore the planet occurrence rate for UCDs within our 40 pc sample.
\end{description}

The 40 pc list of late-type targets and the SPECULOOS target list are publicly available, along with a detailed description of the target selection and survey strategy\cite{Sebastian2021}. On average, we continuously observe one or two targets per telescope. Each telescope operates independently in robotic mode following the plans written by \texttt{SPOCK}. Our strategy is to observe all targets, reaching an effective phase coverage of up to 80\% for temperate planets in Programs 2 and 3 and for planets in the habitable zone in Program 1, resulting in monitoring duration with our SPECULOOS telescope network of 100 to 200 h, respectively.

\section{Data reduction and analysis}
\label{sec:data_reduc}
After the end of the local night, for the case of SSO, science and calibration images are transferred to the online ESO archive. Later these images are transferred to an archive at the University of Liège and a server at the University of Cambridge (UK). The raw data from the SNO and SAINT-EX are directly transferred to the same Cambridge archive. The images transferred to the Cambridge archive are processed by the SPECULOOS reduction pipeline\cite{Murray2020}. The data from SSO and SNO are also processed by the \texttt{prose}\footnote{\url{https://github.com/lgrcia/prose}} pipeline\cite{2022_prose} as an independent set of reduced data. Both pipelines perform image reduction routines (bias, dark and flat-field correction) and use differential photometry algorithm to produce the light-curves. The resulting light-curves reach mmag to sub-mmag precisions with median at 1.5 mmag, and down to 0.26 mmag for the brighest objects with a typical exposure time of 25 s\cite{Sebastian2020DevelopmentProject,Murray2020,Burdanov2022}. SAINT-EX uses a different custom pipeline called PRINCE\cite{Demory2020} (Photometric Reduction and In-depth Nightly Curve Exploration). PRINCE corrects systematics in the light curves through two separate methods. On one side, it uses simple differential photometry, correcting the star's light curve by the median light curve of all stars in the field except for the target star. The second approach is a weighted PCA, removing outliers through an iterative sigma-clipping procedure, where outlier data points are removed, and stars with a large portion of outliers or that appear to be blended or have close neighbours are flagged. In the PCA, each star's data points are weighted by the SNR, with flagged stars, the target, and outliers set to zero weight. The PCA is rerun with uncertainties scaled to achieve a reduced chi-squared of unity for each star's light curve, increasing the weight of well-behaved stars.

Additionally, in order to fully exploit the potential of our data set, we initiated an automatic search for moving objects. We developed a data processing pipeline using  the GPU-accelerated \texttt{tycho tracker} software\footnote{\url{https://www.tycho-tracker.com}}, with two versions; one for daily processing of observations to identify moving objects candidates for the last-night data, and another for analysing archival images to detect slowly moving small bodies. This allows timely submission of new observation and follow-up objects if interest. This pipeline was tested on data from SPECULOOS and their first results and prospect presented in Ref. \citenum{Burdanov2023}.

\begin{figure}[h]
    \centering
    \includegraphics[width=.8\textwidth]{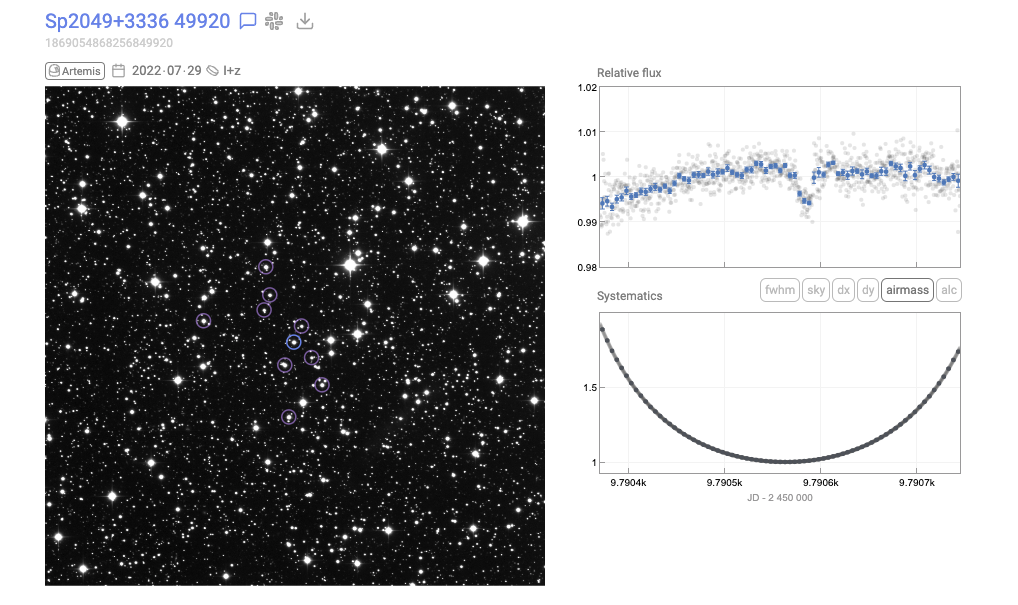}
    \rule[-1ex]{0pt}{4.5ex}
    \caption{PORTAL's web interface. Observation view for a particular night for SPECULOOS-3 target observed by Artemis telescope at SNO. The transit of SPECULOOS-3b planet was detected thanks to PORTAL observation view by our team. The planet was later confirmed by a follow-up campaign by the SPECULOOS team.}
    \label{fig:SPC3_portal}
\end{figure}

The data processed at Cambridge archive is available at the SPECULOOS PORTAL (Pipeline Output inteRacTion Analysis Layer) the afternoon after the observations. These reduced light curves are visually inspected by our team to detect single transit event; this daily inspection allows us to spot any technical problem during observations or to initiate follow-up observations of targets of interest. The recent discovery of SPECULOOS-3b  planet\cite{Gillon2024} was triggered by the visual detection of a single transit in PORTAL (see Fig. \ref{fig:SPC3_portal}), followed by its confirmation after follow-up campaigns (see Sec. \ref{sec:results}). PORTAL also allows members of the SPECULOOS consortium to download and analyse any of the pipeline's outputs and the corresponding technical logs and metadata (further details in Ref. \citenum{Sebastian2020DevelopmentProject}).

\section{Latest developments}
\label{sec:devs}

\subsection{SPIRIT: infrared camera}
\label{sec:spirit}

To improve the precision of observing late-M and L-type stars, we implemented a new near-infrared sensitive instrument to SSO, SPeculoos InfraRed Imager for Transits (SPIRIT\cite{Pedersen2024}). SPIRIT is an InGaAs CMOS-based camera (Princeton Infrared Technologies 1280SciCam, 1280$\times$1024 pixels, 12~\unit{\um} pitch), liquid cooled to $-60$~\si{\celsius}, coupled with a custom wide-pass filter (0.81 – 1.33~\unit{\um}, \textit{zYJ}).

Its wide bandpass was designed to minimise the negative effects of atmospheric precipitable water vapour (PWV)\cite{Pedersen2022} variability on differential photometry whilst maximising flux of its targets of interest. From first light results, the instrument met our expectation on photometric precision and low susceptibility to PWV variability. From our photometric modelling, SPIRIT has the potential to deliver better photometric precision than the existing instrumentation for targets cooler than TRAPPIST-1, 2550~K, primarily limited by readout noise. For robotic applications, it proved to be a more suitable alternative to the traditionally utilised HgCdTe based detectors due to its lower cooling requirements to achieve low dark currents. Further details regarding instrumentation and performance including the first light data analysis is available in Ref. \citenum{Pedersen2024}. 

\subsection{Astra}
\label{sec:astra}

As we transition towards a fully robotic observatory (Pedersen et al., in prep., 2024), we have developed a near-drop-in replacement for the ACP Observatory Control Software and its associated programs (Maxim DL, PinPoint, FocusMax). Our new software, Automated Survey observaTory Robotised with Alpaca (Astra), is a soon-to-be open-source observatory control software designed specifically for automating and managing the operations of survey-focused observatories. Built to seamlessly integrate with ASCOM Alpaca, Astra addresses several limitations of ACP, including the lack of remote heartbeat monitoring and the requirement for a VPN to initiate or terminate observations remotely.

Astra's underlying logic is built with Python, incorporating features such as built-in guiding\cite{McCormac2013}, Gaia-based pointing correction\cite{Gaia,Gaia2018,gaia:2021_EDR3,2022_prose}\footnote{Twirl: \url{https://github.com/lgrcia/twirl}}, auto-focus, pausing observations during unsafe weather, multi-instrument control, and compatibility with Linux, MacOS, and Windows platforms. Since December 2023, we have successfully tested a beta version of Astra at SSO Callisto, and following further development, we plan to deploy it to the remaining SPECULOOS network.

\subsection{Weather station upgrade for SSO}
\label{sec:weather station}
\begin{figure}[h]
\centering
\begin{subfigure}{.5\textwidth}
\centering
  \includegraphics[width=.99\linewidth]{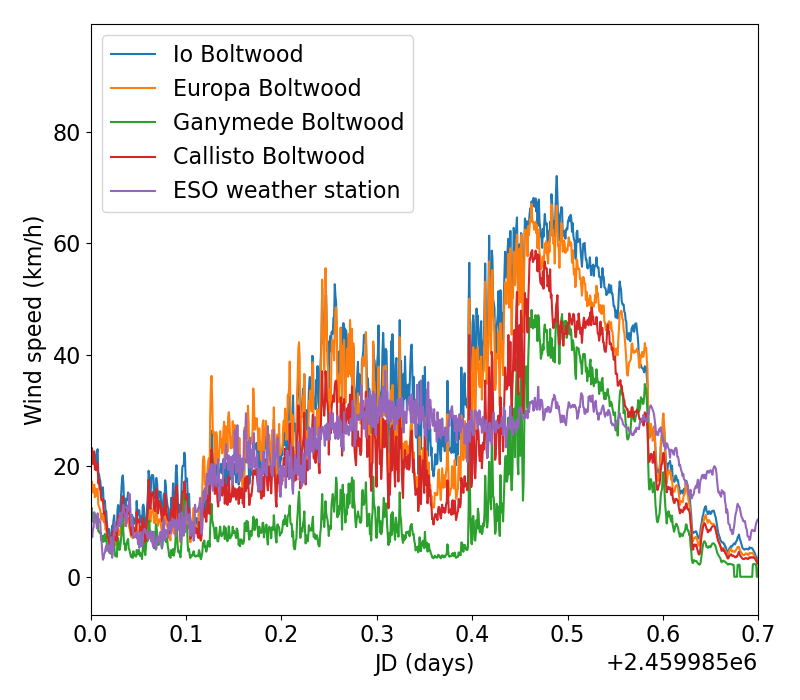}
  \caption{}
  \label{fig:wind_plot}
\end{subfigure}%
\begin{subfigure}{.5\textwidth}
\centering
  \includegraphics[width=.99\linewidth]{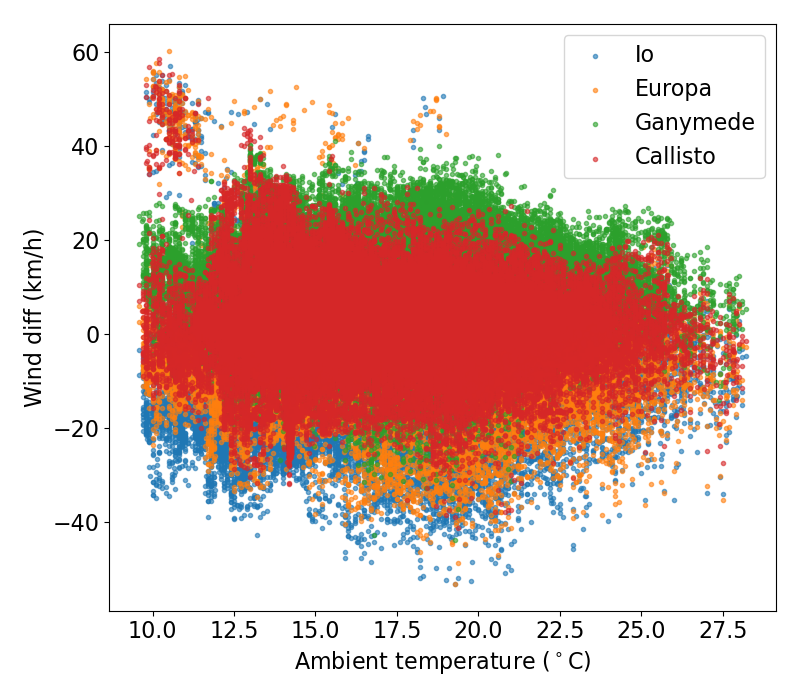}
  \caption{}
  \label{fig:wind_diff}
\end{subfigure}
\rule[-1ex]{0pt}{0.1ex}
\caption{(a): Wind speed measurements from VLT weather station and our Boltwood weather stations. (b): Difference between the measurements of each weather station at SSO telescopes with the VLT weather station versus the measured ambient temperature. The wind speed difference can reach $\sim$40 km/h in some cases.}
\end{figure}

In general, the downtime we have due to adverse weather conditions is within 25-30\% of the available astronomical night per year for all our observatories, however each site has its particularities such as the effect of \textit{calima} in SNO\cite{Burdanov2022}. Given the excellent conditions at Paranal, dust or clouds do not represent a large percentage of downtime at SSO but rather the wind. According to the manufacturer's specifications, our telescopes can operate under safe conditions with winds up to 56 \unit{{\km\per\hour}}. Since our operations are robotic and we do not have a human presence on site at night, we set a more conservative limit on wind speed at 50 \unit{{\km\per\hour}}. Each telescope in SSO has its own weather station (Boltwood cloud sensor 2\footnote{\url{https://diffractionlimited.com/product/boltwood-cloud-sensor-ii/}}) with which the dome closure is activated under adverse weather conditions (for more details see\cite{Sebastian2020DevelopmentProject}). All measurements are consistent between the different weather stations, except for wind speed, which can differ by up to 40 \unit{{\km\per\hour}} in some cases (see Fig. \ref{fig:wind_plot}). As an external point of comparison, we used the data from the weather station on the ESO VLT platform\footnote{\url{https://archive.eso.org/wdb/wdb/asm/meteo_paranal/form}} and analysed the differences relative to the measurements from our weather stations (data from January to March 2023). In this comparison, we did not find any evident relation with ambient temperature (see Fig. \ref{fig:wind_diff}).

To homogenise our wind speed measurements we have used the VLT weather station measurements as a reference to adjust the offset between the weather station of each of our telescopes. Due to the large dispersion of the data, we use just the data from the ESO weather station in the range of $30 - 70$ \unit{{\km\per\hour}} (i.e. $\pm$20 \unit{{\km\per\hour}} around the safe limit value of 50 \unit{{\km\per\hour}}). To robustly adjust the relationship between our weather station data and the ESO one we use Random sample consensus (RANSAC\cite{Fishler1981}) linear regression algorithm implemented in Python. Using the linear regression results, we obtained the wind speed value for each of our weather stations that corresponds to the 50 km/h safety threshold. As a comparison measure we take the results for Io weather station and then we perform a linear regression again but now using the Io data as a reference for the other weather stations. Finally we adopted as a threshold for unsafe wind speed condition the mean value between both configuration results truncated to the integer (Table \ref{tab:weather_station_limit}). Currently these values are the one that we are using in our operation.

\begin{figure}[h!]
\centering
\begin{subfigure}{.5\textwidth}
\centering
  \includegraphics[width=.99\linewidth]{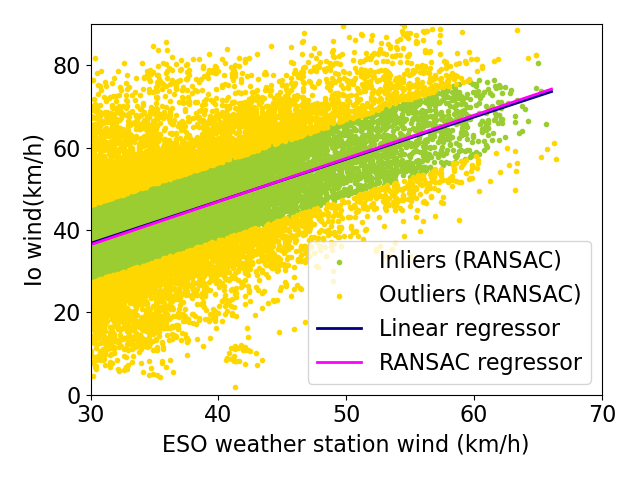}
  \caption{}
  \label{fig:model_io}
\end{subfigure}%
\begin{subfigure}{.5\textwidth}
\centering
  \includegraphics[width=.99\linewidth]{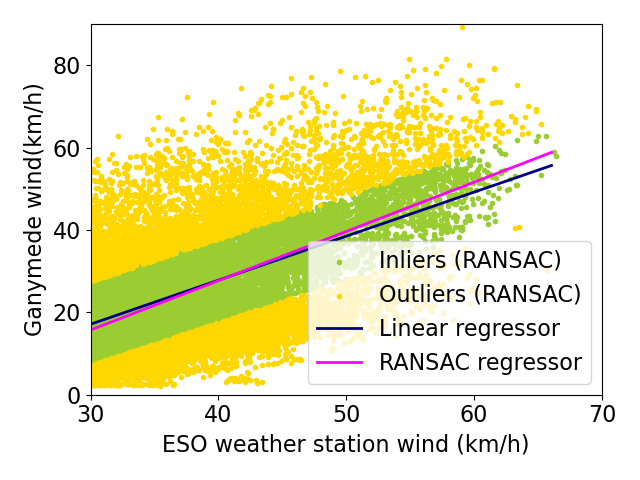}
  \caption{}
  \label{fig:model_gany}
\end{subfigure}
\rule[-1ex]{0pt}{0.01ex}
\caption{Linear fit using ESO weather station as a reference for Io and Ganymede weather station data. We did the same linear regression for Europa and Callisto weather station (see Table \ref{tab:weather_station_limit}).}
\end{figure}

\begin{table}[h]
    \caption{Wind speed values that correspond to the 50 \unit{{\km\per\hour}} safe limit for each of our weather stations depending on calibration source. The last column correspond to the value finally used as threshold for unsafe conditions.}
    \centering
    \vspace{3pt}
    \begin{tabular}{|l|c|c|c|}
    \hline
    \rule[-1ex]{0pt}{3.5ex}  Weather station &   Using ESO weather station & Using Io weather station & Value adopted \\
    \hline
    \rule[-1ex]{0pt}{3.5ex} Io  & 57.6 \unit{{\km\per\hour}} & $\cdots$ & 57 \unit{{\km\per\hour}} \\
    \hline
    \rule[-1ex]{0pt}{3.5ex} Europa & 45.3 \unit{{\km\per\hour}} & 45.4 \unit{{\km\per\hour}} & 45 \unit{{\km\per\hour}}\\
    \hline
    \rule[-1ex]{0pt}{3.5ex} Ganymede & 39.8 \unit{{\km\per\hour}} &  34.3 \unit{{\km\per\hour}} & 37 \unit{{\km\per\hour}} \\
    \hline
    \rule[-1ex]{0pt}{3.5ex}Callisto & 44.4 \unit{{\km\per\hour}}  & 42.4 \unit{{\km\per\hour}} & 43 \unit{{\km\per\hour}}\\
    \hline
    \end{tabular}
    
    \label{tab:weather_station_limit}
\end{table}

The goal of this adjustment was to homogenize our wind speed measurements relative to each other. However, ideally, we would like to have a reliable source of wind speed measurement on-site, especially considering that the VLT weather station is located on the VLT platform. Given the geography of Paranal, the wind conditions at the VLT platform may not necessarily be the same as those where the SPECULOOS telescopes are situated. In this regard, we decided to purchase a Vaisala WXT530 station to be installed at the SSO. In a first stage, the new weather station will be used as a calibration reference for the Boltwood weather station on each telescope. In a second stage the new weather station will be integrated directly in our operation, this development will be in parallel to Astra observatory control software so the integration can be coordinated.

\subsection{Low Earth Orbit satellites detection}
\label{sec:LEO}
\begin{figure}[h]
    \centering
    \includegraphics[width=.5\textwidth]{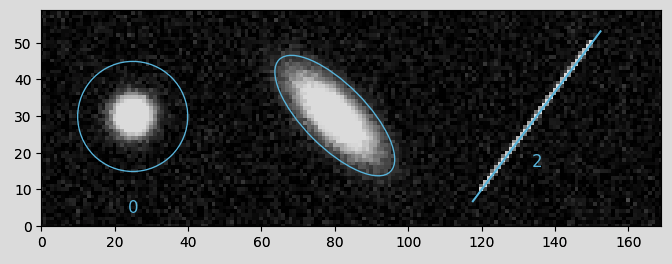}
    \rule[-1ex]{0pt}{4.5ex}
    \caption{Example of  different kind of sources detected by \texttt{prose}. From left to right: Point Source, Extended Source and Trace Source respectively.}
    \label{fig:prose_trace}
\end{figure}

For Low Earth Orbit (LEO) satellites trace detection we use \texttt{prose}\cite{2022_prose}, a Python package to build image processing pipelines for Astronomy. Beyond featuring the blocks to build pipelines from scratch, it provides pre-implemented ones to perform common tasks such as automated calibration, reduction and photometry. In \texttt{prose}, objects detected in astronomical images are represented by Source objects. Where \texttt{prose} features three kinds of sources: \texttt{PointSource}, \texttt{ExtendedSource} and \texttt{TraceSource} (see Figure \ref{fig:prose_trace}).

\begin{figure}[h]
\centering
\begin{subfigure}{.5\textwidth}
\centering
  \includegraphics[width=.85\linewidth]{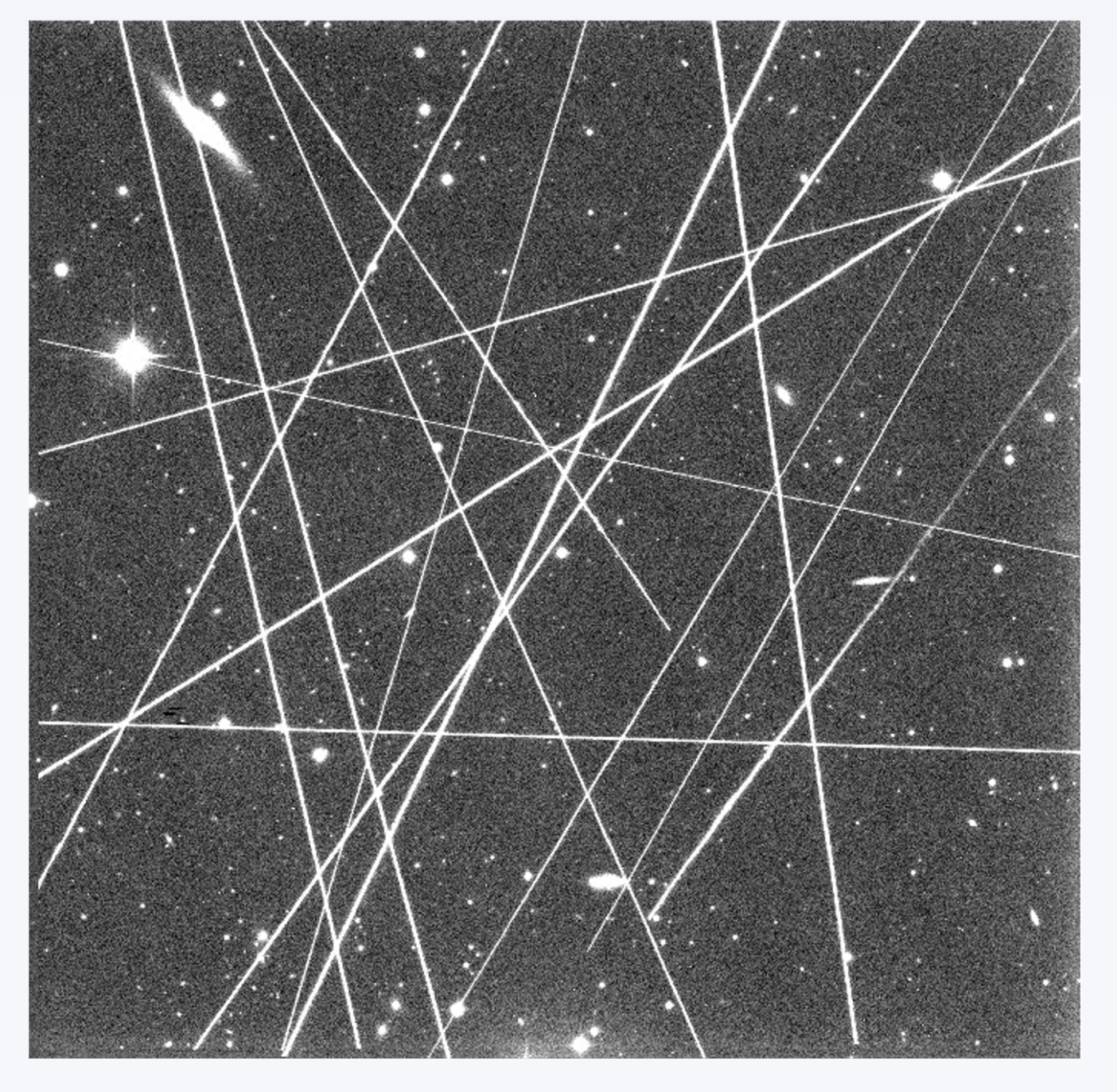}
  \caption{}
  \label{fig:SNO_field}
\end{subfigure}%
\begin{subfigure}{.5\textwidth}
\centering
  \includegraphics[width=.9\linewidth]{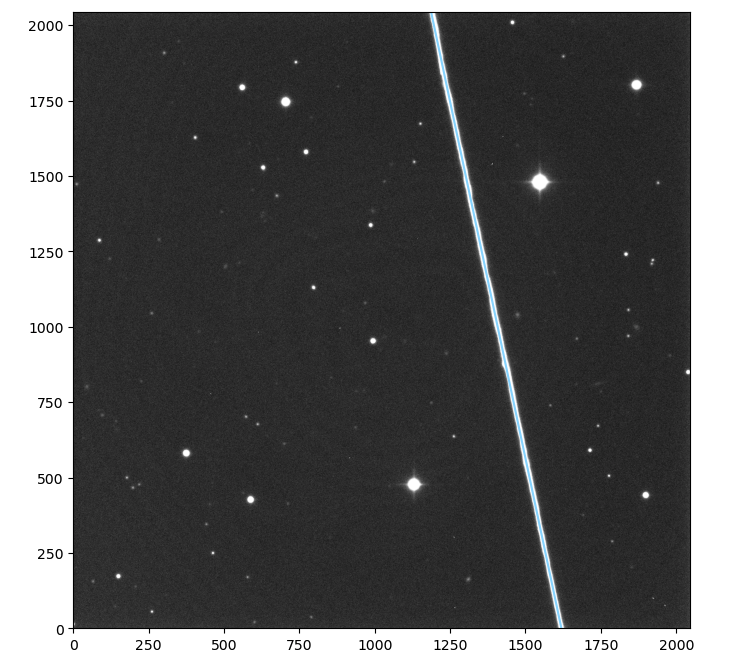}
  \caption{}
  \label{fig:sat_trail}
\end{subfigure}
\rule[-1ex]{0pt}{0.5ex}
\caption{(a): Stack of all images from different nights with satellites traces from one of the SNO fields. (b): SPECULOOS image with satellite trace detected by prose (UTC 2020-11-24T00:31:47.130, exposure time 19 sec).}
\end{figure}

For each image for which we detect a potential satellite trace source (see Figure \ref{fig:SNO_field}), we save the position, orientation and maximum flux of the trace. Additionally we save the julian date of the detection, the image center RA and DEC position, filter, exposure time and FoV of the observation. This information will be helpful for later try to identify the satellite that produced the streak.

To identify Starlink satellite trails on SPECULOOS images, we use a database of Starlink orbital elements encoded in the two-line element sets (TLEs) mainly from CelesTrak\footnote{\url{https://celestrak.com/NORAD/elements/starlink.txt}}. Our code first computes the satellite position in the sky (azimuth, altitude, RA, DEC) and take the trace position from prose detection to compute the angular distance between the sources. In Figure \ref{fig:sat_trail}, we detect a satellite within 10 degrees of the telescope pointing direction. This is a satellite with NORAD number 45406, which corresponds to Starlink-1265 (which was launched on 18-MAR-2020). Given that the field of view of SPECULOOS images is 12x12 arcmin, we cannot confirm that the trace corresponds to that Starlink satellite. Currently we are working on the method to evaluate a reliable identification.

The satellites traces are treated as an elongated ellipsis by \texttt{prose} which lead to wrong vertices points. We are working on an update of this module to obtain proper vertices of each satellites traces in our images. Nowadays we have an updated database with TLEs data for Starlink satellites. We are planning to expand our database to identify other satellites or satellite launcher (currently updating Oneweb TLEs). The next step would be to run our detection and identification codes for all of our data from 2019-2023. We are discussing the better way to store and analyse these data to produce valuable statistics  as well as to produce useful information  for the community, specially for the sathub working group of the IAU Centre for the Protection of the Dark and Quiet Sky from Satellite Constellation Interference (IAU-CPS\footnote{\url{https://cps.iau.org/}}). The goal of this project is try to asses the future (and present) impact not only for our project but also for Paranal site in collaboration with the Atmospheric Scientist of ESO Paranal.

\section{Results and discussion}
\label{sec:results}
SPECULOOS has started observation for $\sim 17\%$ objects of our target list, and for $\sim 40\%$ of them, more than 100h of photometric data have been collected. For $\sim 23\%$ of the program 1 targets the observations have been completed. In parallel, the available high S/N TESS light curves of the brightest targets in our programs are being analysed. Given this synergy, we expect to complete our most time intensive program 1 targets in less than three years.

Beside conducting observations of the SPECULOOS core programs, we dedicate 20\% of our observing capability to annex programs focused mainly, but not limited to, follow-up observations of planet candidates around late-type dwarfs. We are contributing to the follow-up community as part of the TESS follow-up Observing Program (TFOP) working group, especially in the sub-group 1 dedicated to seeing-limited photometry. Our participation in the TFOP working group has brought interesting planet validation such as the large sub-neptune TOI-2406b\cite{Wells2021} or the highly eccentric long-period sub-Neptune TOI-2257b\cite{Schanche2021}.  One notable example is the validation of TOI-715b, a 1.55 $R_{\odot}$ habitable-zone planet. Additionally, we reveal a second candidate planet in this system, TIC 271971130.02, just inside the outer boundary of the habitable zone, and near a 4:3 orbital period commensurability\cite{Dransfield2023}. The follow-up campaign of the TESS transit candidate TOI-1266 was the first exoplanet validation from the SAINT-EX observatory, which revealed a system that hosts a super-Earth and a sub-Neptune around a M3 dwarf\cite{Demory2020}. Recently we reported the discovery of TOI-4336~Ab, an extremely promising target for the detailed atmospheric characterization of a temperate sub-Neptune by transit transmission spectroscopy with JWST. Additionally, we performed follow-up photometry for projects such as NGTS \cite{Guenther2018,West2019}, WASP \cite{Barkaoui2019_wasp,Temple2019} and CHEOPS \cite{Leleu2021}. A recent example is WASP-193b\cite{Barkaoui2024}. Its extended low-density atmosphere makes it an ideal target for characterization through transmission spectroscopy. A single JWST transit observation could provide detailed insights into its atmospheric properties and planetary mass, helping to understand its exceptionally low density and the diverse nature of giant planets.

Giant planets are still a rarity around cool stars, making their detection an important step towards the understanding of their formation pathway. We developed a joint program between TRAPPIST and SPECULOOS that aims to validate photometrically large candidates around M-dwarf stars\cite{Triaud2023}, named M-dwarfs Accompanied by Nearby Giant Orbiters (MANGOs, Dransfield et al. in prep.).

In 2022, we discovered a new potentially habitable planet around LP 890-9\cite{Delrez2022}. The first planet, LP 890-9b (or TOI-4306b), the innermost in the system, was initially identified by TESS. The follow-up observations of LP 890-9 obtained by SPECULOOS have proved fruitful, as they have not only helped to confirm the first planet but have also made it possible to detect a second, previously unknown one. This second planet, LP 890-9c (renamed SPECULOOS-2c by our team, as the star is part of our target list), is similar in size to the first one (about 40\% larger than the Earth) but has a longer orbital period of about 8.5 days. SPECULOOS-2c is one of the most favourable habitable-zone planet for atmospheric characterisation found so far after the TRAPPIST-1 (or SPECULOOS-1) planets. 

More recently, the project discovered SPECULOOS-3b, an Earth-sized planet in a 17-hour orbit around the SPECULOOS target Sp2049$+$ 3336~49920, an M6.5-type ultracool dwarf located 16.8 parsecs away\cite{Gillon2024}. The planet’s high irradiation (16 times that of Earth) combined with its host star's infrared luminosity (K=10.5) and Jupiter-like size (R=0.12 Rsun) make it one of the most promising extrasolar rocky planets for a detailed characterisation by emission spectroscopy with JWST.

\appendix    

\acknowledgments 
 
The ULiege’s contribution to SPECULOOS has received funding from the European Research Council under the European Union’s Seventh Framework Programme (FP/2007-2013; grant Agreement no. 336480/SPECULOOS), from the Balzan Prize and Francqui Foundations, from the Belgian Scientific Research Foundation (F.R.S.-FNRS; grant no. T.0109.20), from the University of Liège, and from the ARC grant for Concerted Research Actions financed by the Wallonia-Brussels Federation. The Cambridge contribution is supported by a grant from the Simons Foundation (PI: Queloz, grant number 327127). The Birmingham contribution research is in part funded by the European Union’s Horizon 2020 research and innovation programme (grant’s agreement no. 803193/BEBOP), from the MERAC foundation, and from the Science and Technology Facilities Council (STFC; grant no. ST/S00193X/1, and ST/W000385/1). J.d.W. and MIT gratefully acknowledge financial support from the Heising-Simons Foundation, Dr. and Mrs. Colin Masson and Dr. Peter A. Gilman for Artemis, the first telescope of the SPECULOOS network situated in Tenerife, Spain. The SPECULOOS North consortium would like to thank IAC telescope operators (Técnico de Operaciones Telescópicas), General and Instrumental maintenance teams for their support on site. TRAPPIST-South is funded by the Belgian National Fund for Scientific Research (FNRS) under the grant PDR T.0120.21. MG is FNRS Research Director. EJ is FNRS Senior Research Associate. This publication benefits from the support of the French Community of Belgium in the context of the FRIA Doctoral Grant awarded to M.T. The postdoctoral fellowship of KB is funded by F.R.S.-FNRS grant T.0109.20 and by the Francqui Foundation. ED acknowledges support from the innovation and research Horizon 2020 program in the context of the Marie Sklodowska-Curie subvention 945298. LD is an F.R.S.-FNRS Postdoctoral Researcher.

\bibliography{report} 
\bibliographystyle{spiebib} 

\end{document}